%% file: ms.tex
\documentclass{aastex}
\usepackage{emulateapj5}

\shorttitle{The nature of 3C~382}
\shortauthors{Gliozzi et al.}

  \def\3c{3C~382}

  \def\feka{Fe K$\alpha$}
  \def\chandra{{\it Chandra}} 
  \def\xmm{{\it XMM-Newton}} 
  \def\suzaku{{\it Suzaku}}
  \def\asca{{\it ASCA}} 
   
  \def\rxte{{\it RXTE}} 
  \def\sax{{\it BeppoSAX}} 
   
  \def\rosat{{\it ROSAT}}

  \def\lum{erg s$^{-1}$}
  \def\flux{erg cm$^{-2}$ s$^{-1}$}
  
  \def\kms{\ifmmode{{\rm km~s^{-1}}}\else{km~s$^{-1}$}\fi}

\def\lsim{\ifmmode{\;\mathop{}^{<}_{\sim}\;}\else{$\;\mathop{}^{<}_{\sim}\;$}\fi}
\def\gsim{\ifmmode{\;\mathop{}^{>}_{\sim}\;}\else{$\;\mathop{}^{>}_{\sim}\;$}\fi}

\begin{document}
\title{The nature of a broad line radio galaxy: Simultaneous RXTE and Chandra HETG observations of 3C~382}

\author{M. Gliozzi}
\affil{George Mason University, 4400 University Drive, Fairfax, VA 22030}

\author{R. M. Sambruna}
\affil{NASA's Goddard Space Flight Center, Code 661, Greenbelt, MD 20771}

\author{M. Eracleous\altaffilmark{1,2}}
\affil{Department of Physics \& Astronomy, Northwestern
University 2131 Tech Drive, Evanston, IL 60208}

\altaffiltext{1}{Permanent Address: The Pennsylvania State University,
Department of Astronomy \& Astrophysics, 525 Davey Lab, University
Park, PA 16802}

\altaffiltext{2}{Center for Gravitational Wave Physics, the
Pennsylvania State University, University
Park, PA 16803}

\author{T. Yaqoob}
\affil{Johns Hopkins University, 3400 N. Charles St., Baltimore, MD 21218}
\affil{NASA's Goddard Space Flight Center, Code 661, Greenbelt, MD 20771}

\begin{abstract}
We present the results from simultaneous \chandra\ and \rxte\ observations
of the X-ray bright Broad-Line Radio Galaxy (BLRG) 3C 382. The long
(120 ks) exposure with \chandra\ HETG allows a detailed study of the
soft X-ray continuum and of the narrow component of the Fe K$\alpha$
line. The \rxte\ PCA data are used to  put an upper limit on the broad line
component and constrain the hard X-ray continuum.  A strong soft excess
below 1 keV is observed in the time-averaged HETG spectrum, which can
be parameterized with a steep power law or a thermal model. The flux
variability at low energies indicates that the origin of the soft
excess cannot be entirely ascribed to the circumnuclear diffuse emission,
detected by \chandra\ on scales of 20--30\arcsec\ (22--33 kpc).  A
narrow ($\sigma<$ 90 eV) Fe K$\alpha$ line (with EW$<$ 100 eV) is 
observed by the \chandra\ HEG. Similar values for the
line parameters are measured by the \rxte\ PCA, suggesting that the
contribution from a broad line component is negligible.  The fact that
the exposure is split into two observations taken three days apart
allows us to investigate the spectral and temporal evolution of the
source on different timescales.  Significant flux variability
associated with spectral changes is observed on timescales of hours
and days. The spectral variability is similar to that observed in
radio-quiet AGN ruling out a jet-dominated origin of the X-rays.  
\end{abstract}


\section{Introduction}
Our understanding of Active Galactic Nuclei
(AGN) is based primarily on radio-quiet (RQ) sources, both at high and
low luminosities. Multi-wavelength observations of Seyfert galaxies
provided us with the
widely accepted view that these sources are powered by accretion of
gas onto a supermassive central black hole, where the emitted light
(optical through X-ray) is produced by an accretion disk and a hot
corona overlaying it (e.g., Haardt \& Maraschi 1991).

On the other hand, not much is known about the central engines of
radio-loud (RL) AGN, due to their relatively low number density (i.e.,
fewer bright examples). X-ray and multi-wavelength studies of broad
line radio galaxies (BLRGs), with $L_{\rm 2-10\;keV}\sim
10^{43}-10^{44}$~\lum, established that these sources exhibit subtle
but significant differences compared to Seyfert 1s.  While their
optical and UV continuum and line emission are similar to those of 
Seyfert 1
galaxies, BLRGs differ from Seyfert galaxies in their X-ray spectral
properties. Specifically, previous \asca, \rxte\, and \sax\
observations of BLRGs showed weaker \feka\ lines and weak or absent
Compton reflection humps at energies $\gsim 10$~keV, a hallmark of
Seyfert 1 galaxies (e.g., Wo\'zniak et al. 1998; Sambruna et al. 1999;
Eracleous et al. 2000; Zdziarski \& Grandi 2001; Hasenkopf, Sambruna,
\& Eracleous 2002; Grandi et al. 2006). These observations, however,
were plagued by limited sensitivity, spectral resolution, and/or the
fact that observations in contiguous bands were not simultaneous 
and were
subject to flux and spectral variability.

The weakness of the \feka\ line and of the Compton reflection are very
important observational clues, since they represent a major difference
between RL and RQ AGN. The origin of this difference, however, is
still debated. The simplest interpretation is that the cold
reprocessor in RL objects subtends a smaller solid angle to the
central X-ray source. This would be the case if the inner accretion
disk was vertically extended as in the ion torus/advection dominated
flow models (Rees et al. 1982; Narayan et al. 1998). In such a
scenario, the RL/RQ dichotomy is caused by different inner accretion
disk structures. Alternatively, the weak
reprocessing features in RL objects might be explained by dilution
effects caused by beamed emission from an unresolved jet.
Another possibility is that the putative hot corona has a
mildly relativistic motion directed away from the disk reducing the
strength of the reflection (Beloborodov 1999; Malzac et al. 2001).
Finally, BLRGs might have more highly ionized
accretion disks than Seyfert 1 galaxies, as a result of higher
accretion rates (e.g., Nayakshin \& Kallman 2001; Ballantyne et
al. 2002). Detailed studies of ionized accretion disk models
(e.g., Matt et al. 1993, 1996; Ross \& Fabian 2005) have demonstrated that
a progressive increase of the ionization parameter, 
$\xi=4\pi F_{\rm x}/n$ (where $F_{\rm x}$ is the X-ray flux and $n$ the 
electron number density), produces several emission lines (e.g., OVII, 
OVIII, and Fe L lines) in the soft energy band as well as a shift
of the energy centroid of the Fe K$\alpha$ line from 6.4 keV to
6.7--6.9 keV. However, when the disk is very strongly ionized
with $\xi>5000 {~\rm erg~cm~s^{-1}}$ , the resulting X-ray spectrum becomes 
virtually featureless, because all the electrons have been stripped off the
atoms. 
In the framework of ionized accretion 
disk models, the weakness of Fe K line can also be explained by
lower values of the ionization parameter ($\xi\sim1000 {~\rm erg~cm~s^{-1}}$) 
provided that
strong relativistic blurring effects are included (e.g., Crummy et al. 2006).
Interestingly, the latter model is also able to self-consistently account for 
the presence of a strong soft excess.

Unfortunately, time-averaged spectra alone are unable to break this
degeneracy, even using broad-band X-ray data with the highest
signal-to-noise (S/N) currently available.  This is illustrated by
recent \xmm\ observations of 3C~120 (Ballantyne et al. 2004; Ogle et
al. 2005) and 3C~111 (Lewis et al. 2005). In the EPIC spectra the
inferred shape of the \feka\ line profile depends sensitively on the
adopted shape of the underlying 0.5--10 keV continuum. The continuum
can be described equally well by very different models, e.g., a simple
power law (yielding broad lines) and a dual absorber (yielding narrow
lines).

Here we discuss in detail simultaneous \rxte\ and \chandra\ HETG
observations of \3c, and our attempt to exploit the complementary
capabilities of these two satellites.  The long (120ks) exposure with
\chandra\ HETG allows a detailed study of the soft X-ray continuum and
of the narrow component of the Fe K$\alpha$ line. The \rxte\ PCA data,
on the other hand, are used to constrain the broad line component as
well as the hard X-ray continuum.
 The fact that the exposure is split into two observations taken three
days apart allows one to investigate the spectral and temporal
evolution of the source on different timescales. Finally, we take
advantage of the unprecedented spatial resolution of \chandra\ to
study the physical conditions of the circum-nuclear region.

3C~382 is a nearby ($z=0.058$), well-studied BLRG with strong and
variable, broad optical lines (FWHM$\sim 11,800~\kms$ for H$\alpha$;
Eracleous \& Halpern 1994). It exhibits a classical Fanaroff-Riley II
radio morphology, with a 1.68\arcmin--long jet extending NE of the
core and two radio lobes, with total extension of 3\arcmin\ (Black et
al. 1992). The inferred inclination angle of the jet of 3C~382 is $i >
15$\arcdeg\ (Eracleous \& Halpern 1998).  The nucleus of 3C~382 is a
bright X-ray source (F$_{\rm 2-10~keV} \sim 3 \times 10^{-11}$
\flux). The 2--10 keV X-ray spectrum is well fitted with a single
power law; when the fit to the hard X-ray spectrum is extrapolated to
lower energies, a strong soft excess is observed (Prieto 2000; Grandi
et al. 2001). \rosat\ observations with the High Resolution Imager
(HRI) revealed extended X-ray emission around \3c (Prieto 2000).
Previous \asca, \sax, and \rxte\ observations showed the presence of a
relatively strong \feka\ line, with $F_{\rm line}\gsim 10^{-5}~{\rm
ph~cm^{-2}~s^{-1}}$ (Wo\'zniak et al. 1998; Sambruna et al. 1999;
Eracleous et al. 2000; Grandi et al. 2001), similar to the Seyfert 1
NGC 5548 studied with the HETG (Yaqoob et al. 2001). The line
equivalent width in 3C~382 ranges from EW=100 eV to EW=700 eV,
depending on the modeling of the underlying continuum.  The limited
sensitivity and resolution of previous X-ray missions prevented
unambiguous modeling of the line and different profiles were derived
by different investigators (as discussed in Wo\'zniak et al. 1998;
Sambruna et al. 1999; and Zdziarski \& Grandi 2001).

The outline of the paper is as follows. In $\S~2$ we describe the
observations and data reduction. The extended circum-nuclear region is
studied in $\S~3$.  The main characteristics of the temporal analysis
are described in $\S~4$. In $\S~5$ we investigate the spectral
properties of the continuum and Fe K$\alpha$ line in \3c.  In $\S~6$
we summarize the main results from the temporal, spectral, and spatial
analyses and discuss their implications.

\section{Observations and Data Reduction}
\3c\ was observed simultaneously with \rxte\ and \chandra\ in October
2004. Both observations were split in two parts: \rxte\ observed \3c\
between October 27 UT 07:16:41 and 28 UT 09:55:15 (net exposure 32.0
ks), and again between October 30 UT 04:28:16 and 31 UT 06:09:15
(exposure 37.8 ks). Similarly, the \chandra\ observations were
performed on October 27 UT 16:50:39 and 28 UT 08:43:45 (exposure 54.2
ks), and between October 30 UT 07:05:20 and 31 UT 01:37:57 (exposure
63.9 ks).

The \rxte\ observations were carried out with the Proportional Counter
Array (PCA; Jahoda et al. 1996), and the High-Energy X-Ray Timing
Experiment (HEXTE; Rotschild et al. 1998) on \rxte. Here we will
consider only PCA data, because the signal-to-noise of the HEXTE data
is too low for a meaningful analysis.  The PCA data were screened
according to the following acceptance criteria: the satellite was out
of the South Atlantic Anomaly (SAA) for at least 30 minutes, the Earth
elevation angle was $\geq 10^{\circ}$, the offset from the nominal
optical position was $\leq 0^{\circ}\!\!.02$, and the parameter
ELECTRON-2 was $\leq 0.1$. The last criterion excludes data with high
particle background rates in the Proportional Counter Units
(PCUs). The PCA background spectra and light curves were determined
using the ${\rm L}7-240$ model developed at the \rxte\ Guest Observer
Facility (GOF) This model is implemented by the program {\tt
pcabackest} v.2.1b and is applicable to ``faint'' sources, i.e., those
with count rates $< 40 {\rm s^{-1}~PCU^{-1}}$. All the above tasks
were carried out with the help of the \verb+REX+ script provided by
the \rxte\ GOF, which calls the relevant programs from the {\tt
FTOOLS} v.5.3.1 software package and also produces response matrices
and effective area curves for the specific time of the
observation. Data were initially extracted with 16~s time resolution
and then re-binned to different bin widths for different applications.
The current temporal analysis is restricted to PCA, STANDARD-2 mode,
2--20 keV, Layer 1 data, because that is where the PCA is best
calibrated and most sensitive. PCUs 0 and 2 were turned on throughout
the monitoring campaign.  However, since the propane layer on PCU0 was
damaged in May 2000, causing a systematic increase of the background,
we conservatively use only PCU2 for our analysis. All quoted count
rates are therefore for one PCU.  We used PCA response matrices and 
effective area curves created
specifically for the individual observations by the program {\tt
pcarsp}, taking into account the evolution of the detector properties.
All the spectra were re-binned so that each bin contained enough
counts for the $\chi^2$ statistic to be valid. Fits were performed in
the energy range 3--15 keV, where the signal-to-noise ratio is the
highest.

The \chandra\ observation was performed with the High Energy
Transmission Grating Spectrometer (HETGS; Markert et al. 1994) in the
focal plane of the High Resolution Mirror Assembly. HETGS consists of
two grating assemblies: a High Energy Grating (HEG; 0.7--10 keV) and a
Medium Energy Grating (MEG; 0.4--10 keV).  The HEG offers the best
spectral resolution in the $\sim 6-7$~keV Fe-K band currently
available ($\sim 39$~eV, or $1860 \rm \ km \ s^{-1}$ FWHM at 6.4 keV).
The MEG spectral resolution is only half that of the HEG. The HEG also
has higher effective area in the Fe-K band.  The HEG and MEG energy
bands are $\sim 0.9-10$~keV and $\sim 0.4-8$~keV respectively, but the
effective area falls off rapidly with energy near both ends of each
bandpass.  The \chandra\ data were reprocessed with version {\tt CIAO}
\footnote{http://cxc.harvard.edu/ciao} version 3.2.1.  and {\tt CALDB} version
3.0.1.  Spectral redistribution matrices ({\tt rmf} files) were made
with the {\tt CIAO} tool {\tt mkgrmf} for each arm ($-1$ and $+1$) for
the first order data of each of the gratings, HEG and MEG.  Telescope
effective area files were made with the {\tt CIAO} script {\tt
fullgarf} which drives the {\tt CIAO} tool {\tt mkgarf}.  Again,
separate files were made for each arm for each grating for the first
order. The effective areas were corrected for the time-dependent
low-energy degradation of the ACIS CCDs using the option available in
the {\tt mkgarf} tool in the stated version of the {\tt CIAO} and {\tt
CALDB} distribution. Events were extracted from the $-1$ and $+1$ arms
of the HEG and MEG using strips of width $\pm 3.6$~arcseconds in the
cross-dispersion direction.  Light curves and spectra were made from
these events and the spectral fitting described later was performed on
first-order spectra that were either combined from the $-1$ and $+1$
orders (using response files combined with appropriate weighting), or
on first-order spectra from the individual $-1$ or $+1$ orders using
the appropriate response files.  The background was not subtracted as
it is negligible in the energy ranges of interest. Examination of the
image of the entire detector and cross-dispersion profiles confirmed
that there were no nearby sources contaminating the data.

The spectral analysis was performed using the {\tt XSPEC v.12.3} 
software package (Arnaud 1996).
The uncertainties on spectral parameters  correspond to the 90\% confidence level
for one parameter of interest ($\Delta\chi^2$ or $\Delta C=2.71$), and
the corresponding luminosities are calculated assuming
$H_0=71{\rm~km~s^{-1}~Mpc^{-1}}$, $\Omega_\Lambda=0.73$ and
$\Omega_{\rm M}=0.27$ (Bennet et al. 2003). With this choice the luminosity
distance of \3c\ is 256 Mpc.

\section{Extended circum-nuclear region}

Previous observations of \3c with the \rosat\ HRI (with spatial
resolution $\sim 4$\arcsec) revealed the presence of extended X-ray
(0.2--2.4 keV) emission around the source, suggesting that the soft
excess of \3c\ is thermal emission of the extended host gas (Prieto
2000). This interpretation was questioned by Grandi et al.  (2001) on
the basis of \sax\ spectral results that are characterized by higher
S/N but encompass a region of 4\arcmin.  Here we use the high spatial
resolution of \chandra\ ACIS-S coupled with its good sensitivity to
clarify this issue.

\begin{figure}     
 \includegraphics[bb=-45 140 650 650,clip=,angle=0,width=12.cm]{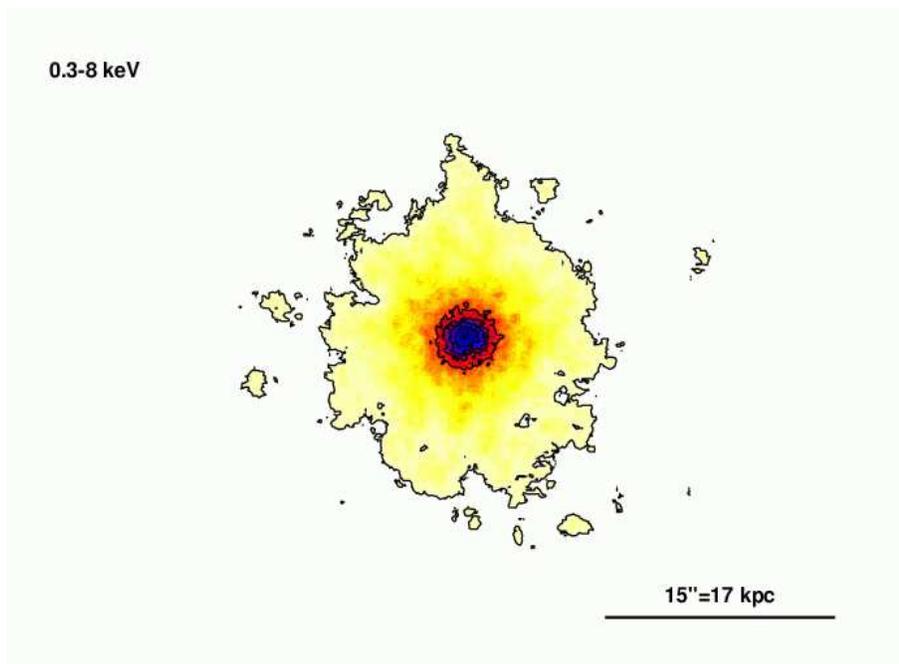}    
\caption{ACIS-S/HETGS zeroth order image of \3c\ in the 0.3--8~keV band.
The image was smoothed using the sub-package {\it fadapt} of     
{\it FTOOLS} with a circular top hat filter of adaptive size, adjusted so as to
include a minimum number of 10 counts under the filter; each final     
pixel is 0{\farcs}1. }  
\label{figure:fig1}   
\end{figure}

Inspection of the 0.3--8 keV image of \3c\ (see
Fig.~\ref{figure:fig1}) confirms the presence of faint diffuse
emission around the core, without any indication for a jet-like
structure.  To investigate the properties of the extended region in a
quantitative way, we extracted a surface-brightness profile from a
series of concentric annuli centered on the position of the central
source. Then, the radial profile was fitted with a model including the
instrument Point Spread Function (PSF). The PSF was created using the
\chandra\ Ray Tracer (\verb+ChaRT+) simulator which takes into account
the spectrum of the source and its location on the CCD (in our case,
we used the best-fitting X-ray continuum summarized in Table~1 and
discussed in {\S}5).

\begin{figure}     
 \includegraphics[bb=70 35 400 515,clip=,angle=0,width=9.cm]{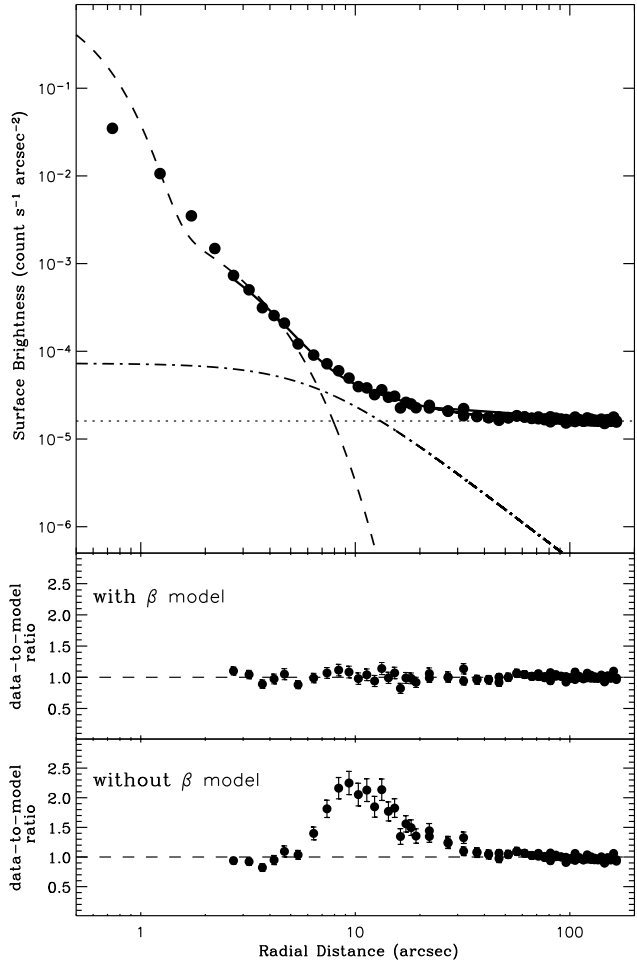}    
\caption{X-ray surface brightness profile of \3c. 
The solid line represents the best-fit model, which comprises     
the PSF model (dashed line), $\beta$--model (dot-dashed line), and     
the background level (dotted line).     
The lower panels show the data-to-model ratios in the cases where a $\beta$-model
is included or not in the fit. The inner region ($r<2${\farcs}5) has been excluded from the fit
because the core is highly piled-up.}   
\label{figure:fig2}      
\end{figure}

The observed radial profile of \3c\ in the total energy band 0.3--8
keV is shown in Figure~\ref{figure:fig2}. The PSF model was normalized
to the second point of the radial profile ($\sim 1${\farcs}5) because
the inner region is affected by photon pile-up and thus the radial
profile is distorted.  Comparing the observed data with the
instrumental PSF (dashed line) plus background (dotted line), excess
X-ray flux over the model is apparent between 6 and 20--30\arcsec\
(see Fig.~\ref{figure:fig2} bottom panel), indicating the presence of
diffuse emission around the core. To model this component, we used a
$\beta$ model, described by the following formula (e.g., Cavaliere \&
Fusco-Femiano 1976):

\begin{equation}     
S(r)=S_0\left(1+{r^2\over r_c^2}\right)^{-3\beta+1/2},      
\end{equation}     

\noindent where $r_c$ is the core radius. The radial profile was then fitted
with a model including the PSF, the background and a
$\beta$-model. The $\beta$-model is required at P$_F \gg$ 99.9\% confidence
according to an $F$-test.  The fitted parameters are: $S_0=
(7.3\pm3.4)\times10^{-5}{\rm ~ct~s^{-1}~arcsec^{-2}}$, $\beta=0.48 \pm
0.05$, $r_c=(6.5 \pm 3.1)$\arcsec, or $\sim$7.2 kpc.  The best-fit
$\beta$ model is plotted in Figure~\ref{figure:fig2} (dot-dashed
line). The middle panel show the data-to-model ratio when a
$\beta$-model is included in the the fit.

We extracted the 0.5--8~keV spectrum of the diffuse emission from an
annular region of inner and outer radii of 3\arcsec\ and 20\arcsec,
respectively.  This spectrum was fitted with a model comprising either
an optically thick or optically thin thermal component (\verb+bbody+
or \verb+bremss+ in \verb+xspec+) and a power-law component, with
Galactic absorption affecting both components. The resulting photon
index of the power-law component is quite low: $\Gamma=0.9\pm0.1$; the
temperature is $kT=0.17\pm0.02$ for the blackbody model, or
$kT=0.45\pm0.10$ if a Bremsstrahlung model is used.  If the soft
component is fitted with a collisionally-ionized plasma model
(\verb+apec+ in \verb+xspec+), at least two components at different
temperatures are required ($kT_1=0.26_{-0.08}^{+0.15}$ keV and
$kT_2=1.2_{-0.3}^{+1.6}$ keV) and the abundances are implausibly low
($Z \ll 0.1\; Z_{\odot}$).  Assuming a thermal Bremsstrahlung model, the
observed flux of the thermal component is $F_{\rm 0.3-2~keV}=4.2
\times 10^{-13}$ \flux\ and the corresponding intrinsic luminosity
$L_{\rm 0.3-2~keV}=6.6 \times 10^{42}$ \lum.  Slightly lower values
are obtained using a black body model for this spectral component.
The power-law component accounts for $\sim 30$\% of the total X-ray
emission in the 0.3--2 keV range. 

The derived luminosity, associated with the extended component in \3c,
is slightly higher than that found in normal elliptical galaxies
(Canizares, Fabbiano, \& Trinchieri 1987), and broadly consistent with
the values found in low-power radio galaxies (Worrall \& Birkinshaw 1994).
The luminosity associated with the power-law component, $L_{\rm 0.3-8~keV}
\sim 1.4\times 10^{43}$\lum, is quite large and  cannot be ascribed
to the integrated luminosities of X-ray binaries in the host galaxies
(see, e.g., Flohic et al. 2006 and references therein). 
Instead, the power-law component might 
be related to the emission from the large-scale jet that is unresolved 
in the X-ray image or to a non-thermal halo already observed in 
in nearby group of galaxies (Fukazawa et al. 2001) and in the X-ray bright 
radio galaxy NGC~6251 (Sambruna et al. 2004).

\section{Temporal Analysis}

The fact that \rxte\ and \chandra\ observations were split into two
parts taken 3 days apart allows us investigation of the temporal and
spectral variability on timescales ranging from few ks to few days.
Between the first and the second exposure, the \rxte\ PCA count rate
decreased from $5.70 \pm 0.02~{\rm s}^{-1}$ to $5.13\pm 0.02~{\rm
s}^{-1}$ in the 2--15 keV energy band. Similarly, the \chandra\ HEG
(1--8 keV) count rate decreased from $0.205\pm 0.002~{\rm s}^{-1}$ to
$0.174\pm 0.001~{\rm s}^{-1}$, and the MEG count rate (0.4--5 keV)
from $0.468\pm 0.002~{\rm s}^{-1}$ to $0.393\pm 0.002~{\rm s}^{-1}$.
In order to check whether the variability shown by  MEG data
is associated with the softest part of the spectrum, we have restricted
the energy band to 0.4--1 keV, which has no overlapping with the 
HEG range. The results ($0.0588\pm 0.0009~{\rm s}^{-1}$ during
the first observation and  $0.0487\pm 0.0007~{\rm s}^{-1}$ during the
second one) indicate that the amplitude of variability is even more
pronounced in the softer energy band. 

This corresponds to a decrease in the average count rate of \3c\ by
factors of 10\%, 15\%, and 17\% in the hard (2--15 keV), medium (1--8
keV), and soft band (0.4--1 keV), respectively, over a time interval of
2~days.

\subsection{The X-ray light curve}

\begin{figure}
\begin{center}
\includegraphics[bb=60 10 440 510,clip=,angle=0,width=10cm]{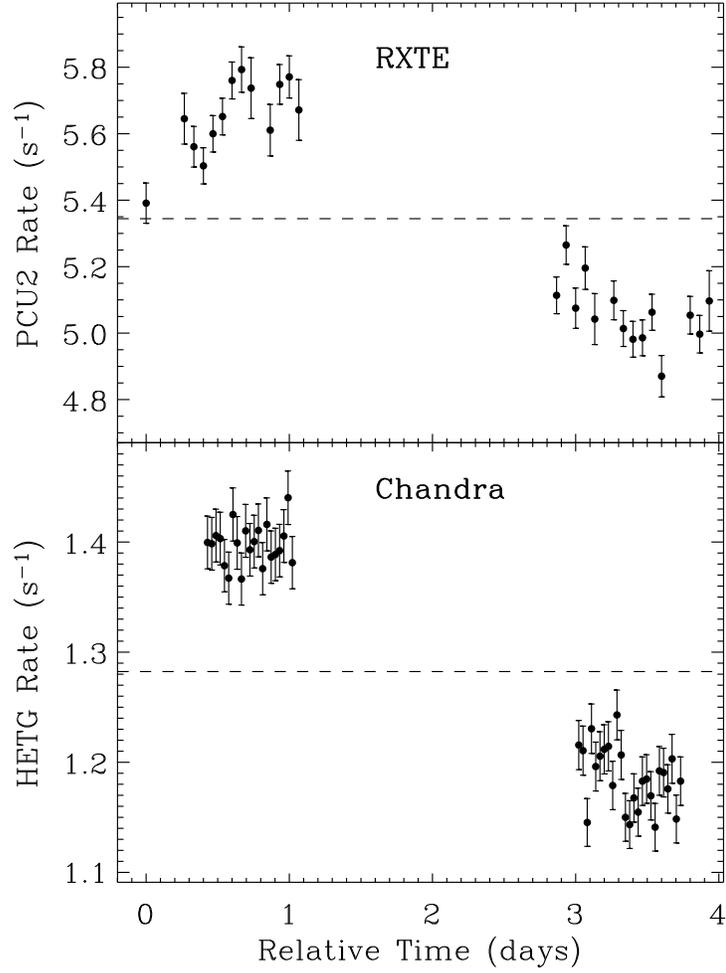}
\end{center}
\caption{Top panel: \rxte\ PCA light curve in the 2--15 keV energy band.
Bottom panel: \chandra\ HETG light curve in the 0.8--7 keV
range. 
The dashed lines represent the respective average count rate.
Time bins are 5760 s ($\sim$ 1 \rxte\ orbit) for \rxte\ and 2560 s for the HETGS light curve.
The time is measured relative to 2004 October 27 UT 07:16:41.}
\label{figure:fig3}
\end{figure}

Figure~\ref{figure:fig3} shows the \rxte\ PCA light curve in the 2--15
keV energy band (top panel) and the \chandra\ HETGS light curve in the
0.8--7 keV range (bottom panel). Time bins are 5760 s ($\sim$ 1 \rxte\
orbit) for \rxte\ and 2560 s for the HETGS light curve.

A visual inspection of Fig.~\ref{figure:fig3} indicates that the \rxte\
and \chandra\ light curves are broadly consistent with each other on
long timescales, which are characterized by an overall decrease of the
count rate. This variability is formally confirmed by a $\chi^2$ test:
$\chi^2=683.9$ for 26 degrees of freedom (hereafter dof) in the case
of \rxte\ and $\chi^2=1019.3$ (45 dof) for the \chandra\ data. Also on
shorter timescales (i.e., within individual exposures), the \rxte\
light curves show significant variability: during the first exposure,
the \rxte\ light curve shows a steady increase of the count rate by a
factor $\sim$10\% ($\chi^2=44.4$, 12 dof), whereas in the second
exposure the \rxte\ count rate steadily decreases by $\sim$10\%
($\chi^2=33.9$, 13 dof). The time elapsed during the first \chandra\
exposure is too short to detect any significant variability. However,
the second \chandra\ observation does show significant short-term
variability: the probability that the count rate is constant 
according to a $\chi^2$ test
is $P_{\chi^2}\sim2\%$ using time bins of 2560 s ($P_{\chi^2}< 1\%$
for $t_{\rm bin}=5760$ s).

\begin{figure}
\begin{center}
\includegraphics[bb=55 50 400 600,clip=,angle=0,width=9cm]{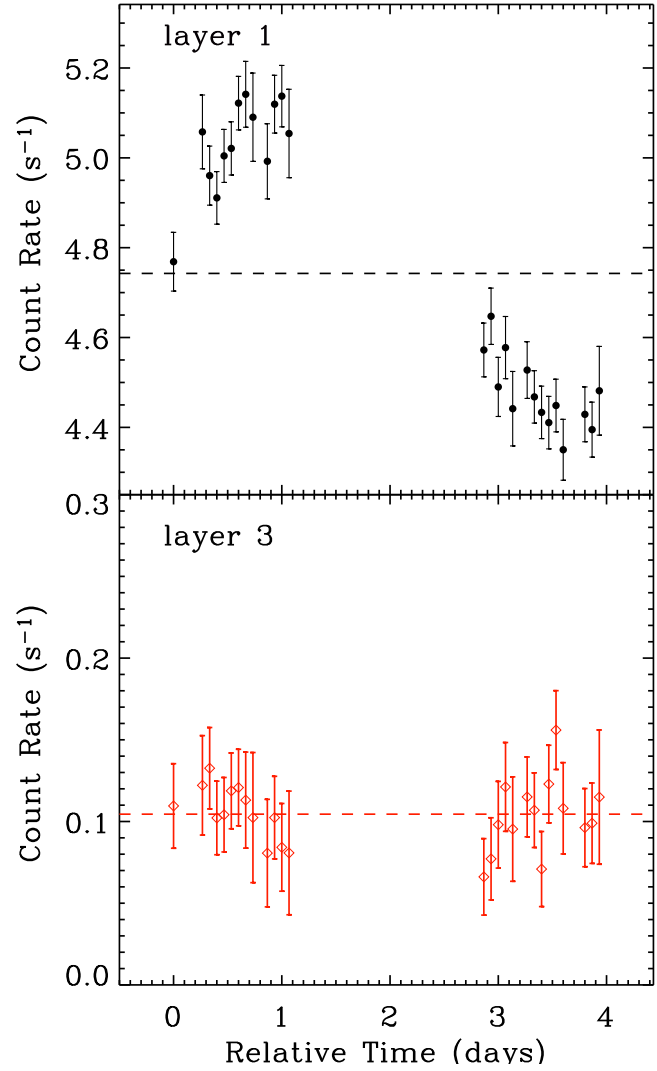}
\end{center}
\caption{Top panel: \rxte\ PCA light curve in the 2--10 keV energy
band, using PCU2 layer 1. Bottom panel: PCU2 layer 3 light curve.
Time bins are 5760 s.  The dashed lines are the average count rate
level. The signal from the source is expected to be very weak in the
layer 3 light curve, thus its variations reflect fluctuations in the
background count rate, Note the difference in scale of the y-axes in
the two plots. The time is measured relative to 2004 October 27 UT
07:16:41.}
\label{figure:fig4}
\end{figure}

Short-term variability may play a significant role in discriminating
between competing spectral models for \3c (see discussion below).
However, the observed variability amplitude is not very large, thus it
is important to verify carefully whether the flux variations are
indeed genuine. In particular, it is necessary to demonstrate that the
variability observed in \3c\ cannot be ascribed to uncertainties in
the \rxte\ background.  To this end, we have performed the following
test: We have compared the background-subtracted light curves obtained
using PCU2 layer 1 and PCU2 layer 3. Since the genuine signal in layer
3 is quite small, its light curve can be used as a proxy to check how
well the background model works. If the latter light curve is
significantly variable with a pattern similar to the one produced
using layer 1, then the variability is simply due to un-modeled
variations of the background. Conversely, if the PCU2 layer 3 light
curve does not show any pronounced variability or if the flux changes
are uncorrelated with those observed in the layer 1 light curve, we
can safely conclude that the short-term variability detected in \3c\
is real.  The two light curves in the 2--10 keV range (where the
background PCA model is better parameterized; see Jahoda et al. 2006
for more details) are shown in Figure~\ref{figure:fig4}, revealing
that the layer 3 time series is consistent with a constant model (both
on long and short timescales) and hence that the variations shown by
layer 1 are genuine.

\subsection{Spectral variability}

In order to investigate whether the flux variability of \3c\ is
associated with spectral variations, we have extracted light curves in
two energy-selected bands and defined the hardness ratio as
$HR=hard/soft$.  For \rxte\ the soft and hard bands are 2--6 keV and
7--15 keV, whereas 0.4--1 keV and 1--8 keV have been chosen for
\chandra.  A $\chi^2$ test of $HR$ versus time indicates that there is
significant spectral variability for \rxte\ ($\chi^2=124.7$, 27 dof)
but not for \chandra\ data ($\chi^2=57.2$, 90 dof).  

\begin{figure}
\begin{center}
\includegraphics[bb=45 30 440 510,clip=,angle=0,width=9.cm]{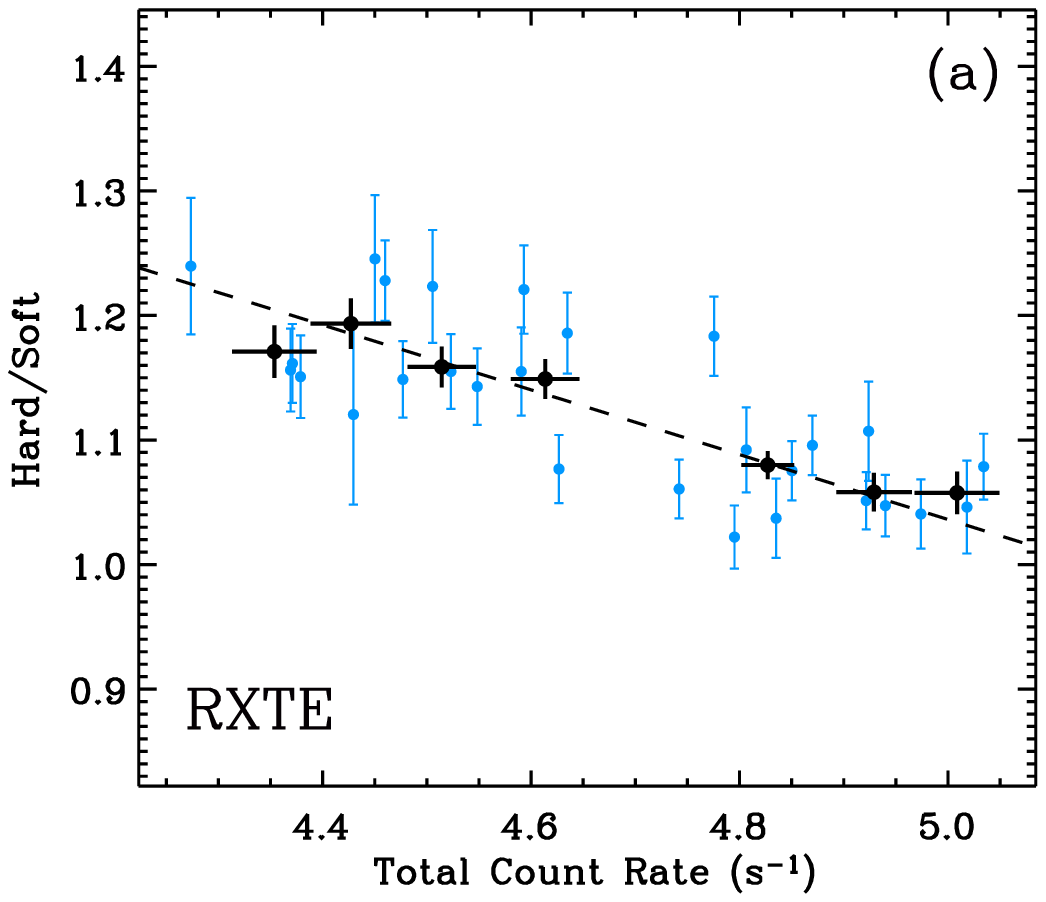}\includegraphics[bb=45 30 440 510,clip=,angle=0,width=9.cm]{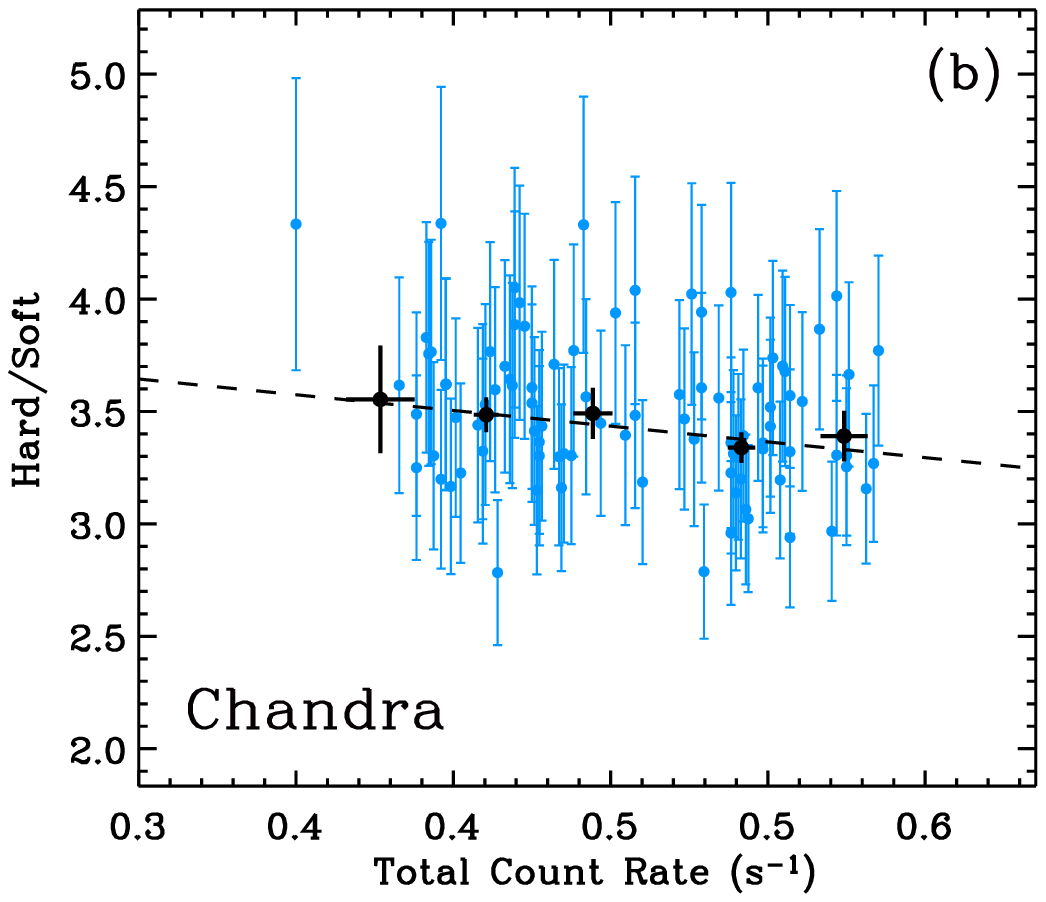}
\caption{(a): Hardness ratio (7--15 keV/2--6 keV) plotted versus the count 
rate for \rxte. The gray (blue in color) filled circles correspond to time-bins
of 5760 s. The black filled circles are binned points obtained taking the 
weighted mean of the original points with fixed bins of 0.1~s$^{-1}$. The 
dashed line represent the best-fit model obtained from a least-squares method.
(b): HR=(1--8 keV/0.4--1 keV) plotted versus the count rate for \chandra\ 
data. Here, the time-bins are 1280 s (gray 
filled circles) and the black circles are the weighted means with fixed bins 
of 0.04~s$^{-1}$.} 
\label{figure:fig5}
\end{center}
\end{figure}

A useful method for investigating the nature of spectral variability
revealed by the hardness ratio  curve is based on the hardness
ratio plotted versus the count rate.  Figure~\ref{figure:fig5}a
shows the hard/soft X-ray color plotted versus the count rate
for \rxte. The gray (blue in color) filled circles correspond to
time-bins of 5760 s. The black filled circles are binned points
obtained taking the weighted mean of the original points with fixed
bins of 0.1~s$^{-1}$.  A visual inspection of Figure~\ref{figure:fig5}a
indicates the presence of a negative trend with the source hardening
when the count rate decreases.  The dashed line represents the
best-fit model to the binned data point, obtained from a least-squares
method: $HR=(2.3\pm0.2) - (0.26\pm0.03)\; r$ (where $r$ is the count
rate).  A similar analysis carried out on \chandra\ data
(Fig.~\ref{figure:fig5}b), suggests the presence of a
similar trend at softer energies, $HR=(4.1\pm0.5) - (1.40\pm0.96)\; r$,
although at a lower significance level.
 
The apparent difference between the \rxte\ and \chandra\ results can 
be mostly ascribed to the larger statistical errors associated with the 
\chandra\ data, a problem which is exacerbated when dealing with the hardness 
ratio. The  different energy bands probed by the two satellites may also play
a role. Indeed, the constant radiation produced by the
extended circum-nuclear component peaks around  0.4--1 keV,
the soft band probed by \chandra. However, the fact that the largest drop
in count rate between observation A and B is measured in the soft band 
(17\% in the 0.4--1 keV range, compared with
15\% in 1--8 keV, and 10\% in 2--15 keV) seems to argue against this 
hypothesis.

The observed spectral trend --the softening of the spectrum that accompanies a
flux increase-- is generally observed in Seyfert-like objects 
(e.g., Papadakis et al. 2002; Markowitz et al. 2003, and references therein) 
and can be explained by two alternative models: 1) a two-component model, 
with two power laws
of fixed photon indices and variable normalization for the softer component
(e.g., Shih et al. 2002), or 2) a single power law with variable photon index
(the ``pivoting model''; Zdziarski et al. 2003). Unfortunately, the shortness 
of the observation and the limited variation of the source's count rate 
hampers 
a more detailed analysis and the possibility of discriminating between the
two competing models. Nevertheless, given the similarity between the spectral
trend shown by \3c\ and the one observed in 3C~120 (a BLRG whose behavior 
has been
interpreted in the framework of the pivoting model; Zdziarski \& Grandi 2001), 
it may be instructive to compute the pivot energy, $E_{\rm p}$, for \3c. 
Following Zdziarski et al. (2003), we obtain $E_{\rm p}\simeq32$ keV.

\begin{figure}
\centering
\includegraphics[bb=40 30 365 300,clip=,angle=0,width=8.cm]{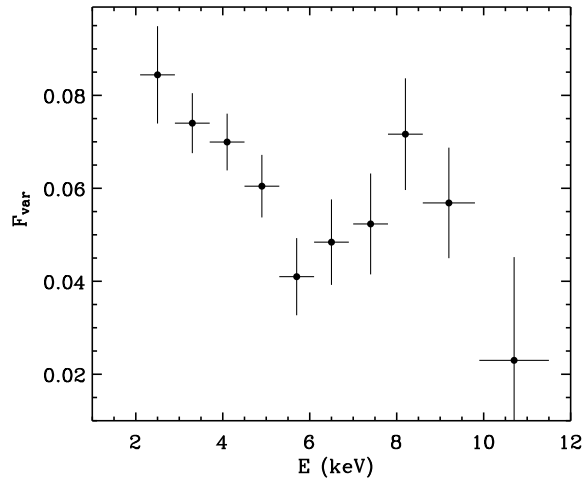}
\caption{Fractional variability parameter plotted versus the
energy for using \rxte\ PCA data. The error-bars along the x axis simply represent
the energy band width. The error-bars along the y axis are computed following
Vaughan et al. 2003. The energy is in the observer rest frame.
} 
\label{figure:fig6}
\end{figure}

Another simple way to quantify the variability properties of \3c,
without considering the time ordering of the values in the light
curves, is based on the fractional variability parameter $F_{\rm var}$
(e.g. Rodriguez-Pascual et al. 1997; Vaughan et al. 2003).  This is a
common measure of the intrinsic variability amplitude relative to the
mean count rate, corrected for the effect of random errors, i.e.,

\begin{equation}
F_{\rm var}={(\sigma^2-\Delta^2)^{1/2}\over\langle r\rangle}
\end{equation}

\noindent where $\sigma^2$ is the variance, $\langle r\rangle$ the
unweighted mean count rate, and $\Delta^2$ the mean square value of
the uncertainties associated with each individual count rate.  We
computed $F_{\rm var}$ on selected energy bands that were chosen to
have similar (and sufficiently high) mean count rates in each band.
(in the end the count rates range between $\sim$0.35 and $\sim$0.70
s$^{-1}$). The plot of $F_{\rm var}$ versus the energy, obtained from
\rxte\ PCA data, is shown in Figure~\ref{figure:fig6}. It indicates
that the amplitude of variability decreases with increasing energy
band, with a minimum around 5.5--6 keV (in the observer frame, where
the Fe K$\alpha$ line is located) and then shows a small bump around
8-9 keV. 
Using \chandra\ MEG data, we have checked the fractional variability
below 1 keV; the value obtained $F_{\rm var,soft}=(6.9\pm1.6) \times
10^{-2}$ is in broad agreement with \rxte\ results in the softer
energy bands.

Since $F_{\rm var}$ is the square root of the excess variance $\sigma^2_{\rm XS}$,
introduced by Nandra et al. (1997) and computed for several Seyfert-like objects, 
we can compare the location of \3c\ in the $\sigma^2_{\rm XS} -  L_{\rm 2-10~keV}$ 
plane. With $\sigma^2_{\rm XS}\sim 3\times10^{-3}$ and 
$L_{\rm 2-10~keV}\sim4.5\times10^{44}$ \lum,  \3c\ would be located in the lower right 
corner of Figure 4 of Nandra et al. (1997), following the same anti-correlation trend
observed in Seyfert galaxies.

The model-independent information provided by $F_{\rm var}$ spectra 
has been frequently used to complement the time-averaged spectral analysis.
For example, Markowitz et al. (2003) carried out a systematic analysis of
the spectral variability properties of Seyfert 1 galaxies observed with
\rxte. Their sample includes objects with both 
broad (e.g., MCG--6--30--15) and narrow (e.g., NGC~4151 and NGC~5548)
Fe K$\alpha$ line profiles. Despite the clear 
difference of the Fe K profiles in the energy spectra,
all objects show  similar trends in the $F_{\rm var} - E$
plane: $F_{\rm var}$ generally decreases with $E$ reaching a minimum around 
6.4 keV and showing a small bump around 8--10 keV. This trend is fully consistent
with the one shown by \3c (see Fig.~\ref{figure:fig6}) and argues in favor of
an accretion-related origin for the bulk of the X-rays. This conclusion is supported
by the results from a similar analysis carried out on \rxte\ observations of 
the blazar Mrk~501. Indeed, this jet-dominated source shows the opposite trend with 
$F_{\rm var}$ monotonically increasing with $E$ (Gliozzi et al. 2006).

Although $F_{\rm var} - E$ plots provide us with a useful tool for distinguishing
between accretion-related and jet-related X-ray emission for AGN, they do not 
help us distinguish between competing theoretical models proposed
to explain the X-ray energy spectra for AGN, such as reflection-dominated
and absorption models (e.g., Gierlinski \& Done
2004, 2006). It must be remarked though that using higher-quality
\xmm\ data for MCG--6--30--15, Ponti et al. (2004) were able to demonstrate that
the relativistically broadened Fe K$\alpha$ line is revealed in the 
$F_{\rm var} - E$ plot by a very
prominent peak in the $\sim$4.5--6 keV energy band, whereas the narrow line 
component,
presumably produced far away from the inner disk, appears as a narrow dip
around 6.4 keV.

In summary, \3c\ shows significant flux variability in all the energy
bands probed on timescales of days and hours. This temporal
variability is associated with spectral variability, with the source
hardening when the count rate decreases. All the variability properties
are consistent with the behavior observed in Seyfert-like objects.

\section{Spectral analysis}

Previous X-ray studies have shown the spectrum of \3c\ to be fairly
complex and provided remarkably different interpretations of its
physical origin. Here, we first investigate the shape of the
continuum, combining data from \chandra\ MEG and HEG, but fitting
separately the \rxte\ PCA data, since the much higher count rate of
the latter would dominate and hence bias the fits.  Then we focus on
the \feka\ line, exploiting the high spectral resolution of the
\chandra\ HETGS and the large collecting area of the \rxte\ PCA.

\subsection{RXTE PCA Continuum}

In the previous section, we found evidence that \3c\ shows spectral
variability between the two exposures. To confirm this finding, we
fitted separately the 3--15~keV PCA spectra from the two observations
with a simple power-law model, modified by Galactic interstellar
absorption.  The results, $\Gamma_{\rm A}$= 1.78$\pm$0.02 during the
first observation (hereafter observation A) and $\Gamma_{\rm
B}$=1.72$\pm$0.02 during the second one (observation B) support the
suggestion that the spectrum hardens as the source's count rate
decreases. 

A simple power-law model (with a Gaussian model parameterizing the
\feka\ line) is a good representation of the \rxte\ spectrum during
observation A,  
but not during observation
B, when the spectrum is best fitted by a broken power law with
$E_{\rm break}\sim$ 8 keV. The results of the spectral fitting of
the PCA data are summarized in Table 1.
During observation A, the observed hard
X-ray flux and the corresponding luminosity are $F_{\rm 2-10
keV}=6.1\times10^{-11}$\flux\ and $L_{\rm 2-10
keV}=4.7\times10^{44}{\rm~erg~s^{-1}}$. A decrease of $\sim$10\%
in the flux and luminosity values is observed during observation B.

Since the presence of an energy  break close to 10 keV accompanied by 
a spectral hardening at higher energies is a classic signature
of Compton reflection, we have substituted the power law with a 
\verb+pexrav+ model 
(which describes the reflection from a neutral disk; see Magdziarz \& 
Zdziarski 1995) in \verb+xspec+, in order to quantify the contribution from
the putative reflection component in \3c.
Due to the short exposure and the limited energy range,
all the pexrav parameters, except the photon index $\Gamma$, the
reflection fraction $R$, and
the normalization, were kept fixed at reasonable values (we adopted the values 
given by Eracleous et al. 2000). 
The spectral fitting suggests that the reflection fraction
is more pronounced during observation B ($R \sim$ 0.6) than during the
first observation ($R \sim$ 0.2). However, the large uncertainties prevent any
firm conclusion; for example, the value of $R$ derived during observation A is
consistent with zero at the 90\% confidence level. 
Statistically, these fits are as good as of those
obtained using power-law or broken power-law models.

We have also tried to substitute \verb+pexrav+ with the \verb+pexriv+ 
model,
which describes the reflection from an ionized disk (Magdziarz \& 
Zdziarski 1995). The resulting  value of the ionization parameter $\xi$ is 
consistent with zero, although completely unconstrained during observation A.
No statistical improvement is obtained with this model.
However, the poor spectral resolution of the PCA coupled with
the limited energy range non-background-dominated (3--15 keV) hampers
the spectral analysis and makes it impossible to constrain the spectral
parameters.

\subsection{Chandra HETG Continuum}
The \rxte\ PCA spectra are characterized by a high S/N and low
spectral resolution, while the opposite is true for the \chandra\
HETGS spectra. Therefore, in order to study the continuum in the
\chandra\ spectrum we combine the positive and negative grating orders
and bin the spectra heavily. The MEG spectra are binned at 0.08~\AA,
whereas the HEG data are binned at 0.04~\AA. \footnote{For reference,
we note that the width of a spectral bin in energy space is related to
its width in wavelength space via $\Delta E = 26\;(\Delta
\lambda/0.08\;{\rm\AA})\;(E/2\;{\rm keV})^2$~eV.} All the spectra are
grouped to contain at least 15 counts per bin for the $\chi^2$
statistic to be valid. Before combining $+1$ and $-1$ orders, we have
checked the spectra from individual arms: Spectra of order $-1$ are
consistently steeper that those obtained with order $+1$, however, the
differences are well within the statistical and systematic errors.

The 2--8 keV HEG spectra confirm the hardening of the source
between the two observations: $\Gamma_{\rm
A}$=1.70$\pm$0.05 and $\Gamma_{\rm B}$=1.63$\pm$0.05. As already
noticed in past studies based on \rxte\ and \chandra\ simultaneous
observations (e.g., Yaqoob et al. 2003), the HEG photon indices appear 
to be flatter by $\lsim 0.1$ compared to those obtained with the PCA;
in any case, the spectral results are consistent within the systematic errors.

\begin{figure}
\includegraphics[bb=60 30 580 720,clip=,angle=-90,width=9.cm]{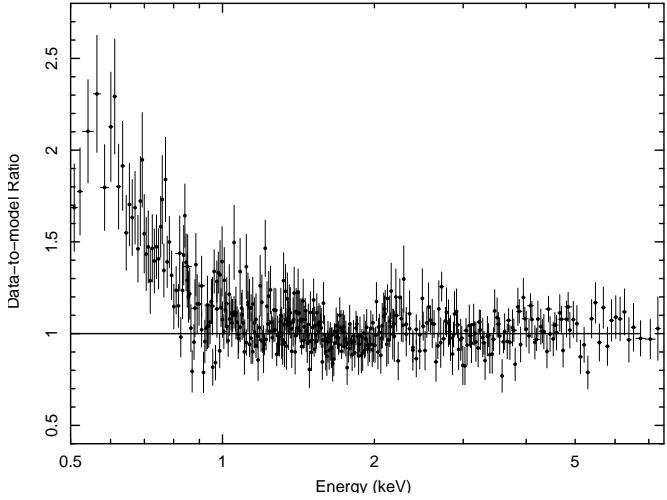}
\caption{Data/model ratio to a simple power-law model modified by
photoelectric absorption in the Milky Way. The data are from the MEG
and HEG during the first observation. A strong soft excess in present
below 1 keV.}
\label{figure:fig7}
\end{figure}

If the best-fit power-law model is extrapolated to softer energies
combining MEG and HEG data, a strong soft excess below 1 keV is
clearly visible (see Fig.~\ref{figure:fig7}).  A very similar result
is obtained if the same procedure is applied to observation B.  This
clearly indicates that more complex spectral models are necessary to
fit also the soft X-ray spectrum of \3c.

\begin{figure}
\includegraphics[bb=40 1 580 720,clip=,angle=-90,width=9.cm]{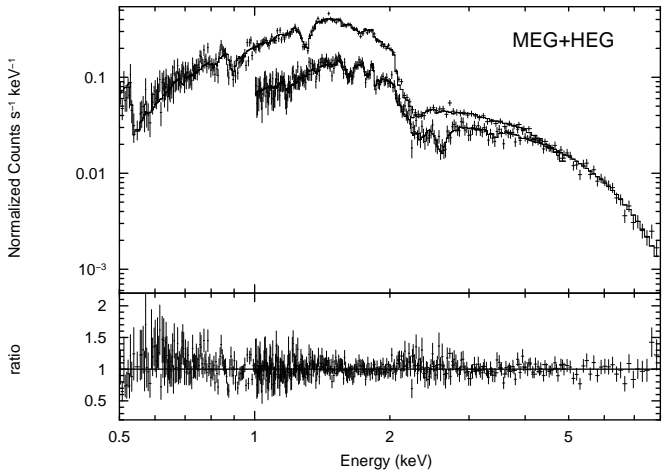}
\caption{The MEG and HEG spectra in the 0.5--8 keV energy range,
obtained combining orders $+1$ and $-1$ during observation A when the
mean count rate was higher (a very similar soft spectrum is obtained
during observation B). The bin size are 0.08~\AA\ and 0.04~\AA\ for the
MEG and HEG, respectively. The
continuous line represents a double power-law model absorbed by
Galactic $N_{\rm H}$. The bottom panel shows the data/model ratio.
}
\label{figure:fig8}
\end{figure}

We fitted the 0.5--8 keV combined MEG and HEG spectra with several
phenomenological models, such as a power law describing the hard
energy spectrum combined with an additional component (steep power law
or thermal component) parameterizing the soft excess.  The results of
the spectral fitting, summarized in Table 2, indicate that all models
are able to fit the broad-band continuum of \3c fairly well. This is
confirmed by Figure~\ref{figure:fig8}, where the MEG 
and  HEG data from 0.5--8 keV  are
shown with the double power-law model superimposed.  
The observed soft
X-ray flux and the corresponding luminosity are very similar for all
models. During observation A, using the double power-law model we 
obtain: $F_{\rm 0.4-2 keV}=3.2\times10^{-11}$\flux\ and $L_{\rm 0.4-2
keV}=3.6\times10^{44}{\rm~erg~s^{-1}}$. The values of the flux and 
luminosity in the 0.4--2 keV band decrease by a factor \gsim 15\%
during observation B.

Despite the formally acceptable fit (see Table 2),
the plot of the soft spectrum with the best-fit model superimposed 
(Fig.~\ref{figure:fig8}) suggests the presence of several line-like features
in the 0.7--1 keV range (in the observer's frame). In order to investigate
further this issue, after restricting the fitting range to 0.5--1.5 keV,
we have tried to add several Gaussians to the underlying continuum. We
find that the fit is improved at the 90\% confidence level (i.e., $\chi^2$
decreases by more than 6.25 for each Gaussian added) by adding two narrow lines.
The energy centroids $E_1=0.89\pm0.01$ keV and $E_2=1.04\pm0.01$ keV (in the
source rest frame) are consistent with transition from Ne IX and Ne X,
respectively. Detailed photoionization modeling may give more insight into
the identification of these lines but there is insufficient
statistically significant information in the spectrum to warrant
more sophisticated modeling.

Similarly to the procedure applied to the PCA data, we used 
the \verb+pexrav+ model
instead of the power law to account for the presence of reflection.  Fixing
$R$ at the best-fit values obtained from the PCA data, the resulting
spectral fits are as good as those obtained using the phenomenological models
reported in Table 2. If $R$ is left free to vary, the spectral fits yield $R=0$
with 90\% upper limits of 0.3 and 0.6 for observation A and B, respectively.
Using the \verb+pexriv+ model with HETG spectra leaves the ionization
parameter totally unconstrained.

As an alternative to the phenomenological models described above, we have also 
tried to fit the HETG spectra with a more physically-motivated model such as
\verb+reflion+  (which describes the reflection from an optically-thick 
atmosphere of constant density; see Ross \& Fabian, 2005)
in \verb+xspec+. However, the low S/N of the data
in the soft part of the spectrum combined with the limited energy band
hampers the analysis and does not allow to constrain the parameters
properly.  The
resulting fit is poor (the soft excess is still present) and the best
fit parameters indicate that the fraction of the reflected radiation
is quite small($F_{\rm refl}/F_{\rm tot}\sim 10\%$)
and also that the ionization parameter is low
($\xi \sim 350-400~{\rm erg~cm~s^{-1}}$).  A
significant formal improvement in the fit is obtained by convolving
the reflection disk model with \verb+kdblur+ to account for the
relativistic blurring close to the black hole (see Crummy et al. 2006
for a detailed description of \verb+kdblur+).
In this case the
reduced $\chi^2$ is comparable to the one obtained with the
phenomenological models, but the number of free parameters is
significantly larger. However, the fitting procedure becomes extremely
slow and the parameters of \verb+kdblur+ (specifically, $r_{\rm in}$,
the disk emissivity index, and the inclination angle) remain totally
unconstrained. 
If this model with  parameters fixed at their best-fit values is
applied to the PCA spectra, the resulting fits are significantly worse
than those obtained with power law or pexrav models. 

Finally, if the
\rxte\ data are fitted simultaneously with the HETG spectra, an adequate
fit of the data 
($\chi^2_{\rm red}=1.09/536$ for observation A and 
$\chi^2_{\rm red}=1.08/538$ for observation) 
is obtained using a power law to parameterize the soft excess
and a pexrav model to describe the spectrum at higher energies. The
resulting deconvolved $E~F_{\rm E}$ spectra are shown in Figure~\ref{figure:fig10}.

In summary, \3c\ shows a strong soft excess below 1 keV, when the hard
(2--10 keV) photon index is extrapolated at softer energies. The
broad-band spectrum is fitted reasonably well by a power law plus a
thermal component (or a steep power law).  The limited quality of the
data hampers the use of more complex spectral models such as ionized disk
reflection models.  No intrinsic absorption in addition to the
Galactic column density is required.

\subsection{Fe K$\alpha$ line}

We use the complementary characteristics of the \chandra\ HEG and the
\rxte\ PCA to study the profile origin of the \feka\ line and
investigate its origin. Because of its high spectral resolution, the
HEG probes the narrow component of the line profile (likely to
originate in matter that is not part of the inner accretion disk).  In
contrast, the PCA is sensitive to the entire \feka\ profile, which may
consist of both the narrow and the broad component (the broad
component is thought to be produced in the inner part of the accretion
disk).

Since we want to test whether the narrow line is resolved by the HEG
(the resolution element is 0.012~\AA), we use spectra binned at
0.01~\AA. To analyze the HEG spectrum, we restrict our fits to the
energy range to 3--8 keV and use a power-law model for the local
continuum. To judge the goodness of the fit, we use the $C$-statistic.
We fit the PCA spectrum in the 3--15 keV range adopting a broken
power-law model for the continuum and using the $\chi^2$ statistic as
an indicator of the goodness of the fit. The profile of the \feka\
line is described by a Gaussian model in all cases.

The results of this analysis are reported in Table 3. The values of the
spectral parameters remain basically unchanged when we use higher 
resolution spectra binned at 0.005~\AA.
 We fitted
spectra from different dispersion arms separately and checked the
results for consistency. When we use the HEG spectrum of order
$+1$ only, the significance of the \feka\ line detection is much
higher, therefore, we report both the results obtained after combining
the $+1$ and $-1$ orders and those from order $+1$.  Using the
combined spectrum from both arms, a weak (EW$\sim 20-50$ eV),
unresolved line with energy consistent with Fe  K$\alpha$ at 6.40
keV is detected at a high confidence level during observation B, but
it is only marginally significant during observation A. As expected,
the spectral parameters are better determined during observation B,
when the continuum level is lower, although there is a substantial
agreement between the line parameters during the two observations.  It
is worth noting that using spectra from dispersion arm +1, the line
significance and strength are substantially increased in both
observations, and that during observation B the line appears to be
resolved ($\sigma=81_{-25}^{+35}$ eV, corresponding to a FWHM of
8900~\kms) at the HEG resolution.

The \rxte\ PCA spectrum, fitted separately, requires a line at $\sim$
6.4 keV (in the source rest frame) in both observations. Importantly,
when fitted with a Gaussian model, the spectral parameters are fully
consistent with those obtained with the HEG (see Table 3). In
particular, taking into account the systematic error in the \rxte\
flux (known to be around 10--20\% higher than \chandra; see Jahoda et
al. 2006), the line intensities measured by the \rxte\ PCA are in
fairly good agreement with those measured by the HEG (see also
Fig.~\ref{figure:fig10}). This suggests that the \feka\ line in \3c\ is
dominated by a narrow component and that a broad component is not
detected. In order to quantitatively constrain the broad line component,
we added to the best-fitting continuum a \verb+diskline+ (which describes 
the profile of a line emitted
from a relativistic accretion disk; the parameters were fixed at the
following values $r_i=6~r_{\rm g}$, $r_o=400~r_{\rm g}$, 
$q=2.5$, $i=30\arcdeg$, and $E=6.4$ keV; where $r_{\rm g}\equiv GM_{BH}/c^2$).
The model also includes a narrow Gaussian line. The addition of a broad line
component does not improve the fit at all, but it allows the determination 
of the 90\% upper limits on the line equivalent width, wich are 40 eV during
observation A and 90 eV during observation B. Similar upper limits are obtained
if we use the \verb+laor+ model, which describes the line emission in the
hypothesis that the black hole is nearly maximally rotating.

It is instructive to compare the measured \feka\ equivalent widths with the
values expected on the basis of the reflection fraction $R=\Omega/2\pi$ obtained 
by fitting the contimuum with a \verb+pexrav+ model. According to the calculations
from George \& Fabian (1991), the relation between these two quantities can be
expressed as $EW=160(\Omega/2\pi)$ eV, which yields $EW\sim$ 40 eV and 
$EW\sim$ 100 eV, for observation A and observation B, respectively. These values, 
which are in general agreement with the measured values of $EW$ (see Table 3), 
suggest a common physical origin for the  \feka\ line and the Compton reflection 
component.

\section{Summary of Results and Discussion} 

By taking advantage of the complementary capabilities of the \rxte\
PCA and the \chandra\ HETGS we have obtained several 
results on the BLRG \3c, which can be summarized as follows:

\begin{itemize}

\item 
A model-independent timing analysis has revealed the existence of
significant flux variability on short (few ks) and medium (days)
timescales. This temporal variability is accompanied by spectral
variability such that the source spectrum hardens as the count rate
decreases. The variability amplitude decreases with increasing
energy. A potentially important clue is that the soft band shows 
significant flux
variability, coordinated with the hard band, with a similar
amplitude. This suggests a close connection between the two energy
bands.

\item
An analysis of the time-averaged spectrum shows the presence of a
strong soft excess below 1 keV, when the hard (2--10 keV) power law is
extrapolated to lower energies. The broad band spectrum is adequately
fitted by a power law plus a thermal model (or a steep power law)
describing the soft excess. The spectra are fitted equally well 
with a neutral reflection model (pexrav). The reflection fraction is
poorly constrained during observation A ($R\sim0.2$ but consistent with zero at
the 90\% confidence level) and is of the order of 0.6 during observation B,
when the continuum flux is lower.

\item

A weak, narrow iron line with energy centroid consistent with neutral
Fe K$\alpha$ is detected by the PCA and the HEG at high confidence
level. There is no indication of a relativistically broadened Fe line.
The good agreement between the line parameters obtained with the PCA
and those yielded by the HEG suggests that the \feka\ line in \3c\ is
dominated by a ``narrow'' component, probably not originating in the
inner accretion disk.  The FWHM of the \feka\ line of 
$8900\pm3500$~\kms\  (as measured by HEG+1 during observation B) is
comparable to the FWHM of the optical hydrogen Balmer lines of
11,800~\kms, suggesting a possible common origin of these lines. The
double-peaked optical lines have been attributed to the outer
accretion disk (Eracleous \& Halpern 1994, 2003), which is also a
plausible site for the production of the \feka\ line. 

\item
A spatial analysis based on the radial surface brightness profile
confirms the presence of diffuse emission around \3c\ on scales of the
order of 20--30\arcsec\ corresponding to $\sim 22$--33 kpc. The soft
spectrum of the diffuse emission is well described by a thermal model
(black body or Bremsstrahlung) with temperatures significantly higher
than those describing the soft excess in the AGN spectrum. This
result, combined with the variability observed in the soft band, rules
out the hypothesis that the soft excess in the AGN spectrum can be
explained entirely by the extended emission.

\end{itemize}

If we compare our spectral results with the most recent X-ray studies
of \3c\ in the literature, i.e.  with \sax\ analysis of Grandi et
al. (2001) and the \rxte\ investigation from Eracleous et al. (2000),
we find a general agreement on the weakness of the iron line:
equivalent widths of 50 and 90 eV were measured by \sax\ and \rxte,
respectively, which are fully consistent with the results reported in
Table 3.  Also the weak reflection component measured in the continuum, 
although poorly constrained, is in broad agreement
with the results reported in the literature ($\Omega/2\pi\sim 0.5$ by Eracleous 
et al. 2000; $\Omega/2\pi\sim 0.3$ by Grandi et al. 2001).

Our analysis of the time-averaged spectrum alone has  not led to definitive
conclusions about the nature of X-ray source in \3c. However, the
combination of the information from the temporal, spectral, and
spatial analyses, can put tighter constraints on \3c\ with
implications for all BLRGs.

Previous studies of time-averaged spectra have shown that several
competing scenarios can explain the weak reprocessing features
observed in the X-ray spectra BLRGs equally well.  In summary, these
peculiar X-ray properties can be explained by: 1) an accretion disk
truncated at small radii with a radiatively inefficient accretion flow
in the central region (RIAF; Eracleous et al. 2000); 2) a highly
ionized accretion disk (Ballantyne et al. 2004); 3) a mildly relativistic
outflowing corona (Beloborodov 1999); and 4) dilution from jet
emission (e.g., Sguera et al. 2005).

To evaluate the above hypotheses we use the black hole mass in \3c,
obtained from the B-band luminosity of the host galaxy,
$M_{BH}=1.1\times10^9{\rm~M_\odot}$ (with an uncertainty of
approximately 40\%; Marchesini et al. 2004). We also estimate the
bolometric luminosity \footnote{We obtain the bolometric luminosity
from the observed 1--2~keV and 1--10~keV luminosities of \3c\ using
the relations $L_{\rm bol}=10\;L_{\rm 1-10\; keV}=80\;L_{\rm 1-2\;
keV}$. The scale factors were found from the spectral energy
distributions of seven radio loud quasars with comparable X-ray
luminosities to \3c, reported by Elvis et al. (1994), namely, 3C~48,
PKS~0312--77, 3C~206, 3C~249.1, 3C~323.1, 3C~351, and 4C~34.47. We
estimate the uncertainty to be about a factor of 2.}  to be $L_{\rm
bol}=9\times 10^{45}~{\rm erg~s}^{-1}$. Thus we obtain an Eddington
ratio of $L_{\rm bol}/L_{\rm Edd}=0.02$--0.2 (including our best
estimate of the uncertainties) and a light crossing time of the inner
accretion disk of $\tau_{\ell} \equiv 100\; r_{\rm g}/c = 2$--5~days. 
With the above estimates in
mind, our results can constrain the proposed scenarios as follows:

\begin{itemize}

\item 

If the inner accretion disk is a radiatively inefficient accretion flow 
(RIAF), then its size (i.e., $r_{\rm
tr}$, the transition radius from the thin disk to the hot, vertically
extended flow) is constrained by the observed variability time
scale. The dramatic drop in the observed X-ray flux over a timescale
of approximately two days, sets a stringent upper limit to the
light-crossing time of the RIAF ($r_{\rm tr} < 100 r_{\rm g}$). Such a
small transition radius is not unprecedented (see, for example the
application to NGC~4258 by Gammie, Narayan, \& Blandford 1999).  There
are also short-term fluctuations with an amplitude of about 5\%
superposed on the secular flux changes during our observations, which
may result from inhomogeneities in the flow. The inferred Eddington ratio is
uncertain enough that its lower limit ($L/L_{\rm Edd}\approx 0.02$) is
compatible with  presence of a RIAF.  For higher values of $L/L_{\rm Edd}$,
an inner radiatively inefficient flow might still be a viable solution
in the form of the luminous hot accretion flow proposed by Yuan et al. (2007).

\item

The scenario in which the X-rays are dominated by reflection from a
highly ionized accretion disk cannot be firmly ruled out or confirmed
by the present data.
The time-averaged spectral analysis reveals no clear evidence
for a high ionization state. However, if the ionization is very high
or if relativistic blurring effects are very strong, no prominent
spectral signatures are expected and  in any case the current data 
would not be able to detect them unambiguously. 
In principle, useful information about this scenario 
can also be obtained from the estimated accretion rate
and from the measured variability. Specifically, all ionized accretion
disk models require very high values of $\dot m$ to produce high ionization
states. However, the uncertainties on the black hole mass and hence on
the Eddington ratio estimated for \3c\ do not allow one to derive firm 
conclusions from this argument.  
Finally, the ionized-reflection-dominated scenario, at least in the framework
of the photon bending model (see Miniutti \& Fabian 2004 and
references therein), predicts that the primary (in our case the 2--10 keV
component) and the reflection-dominated components (the soft
excess) should not vary in concert, when the source is in a  bright state.
Therefore, the simultaneous flux drop observed for \3c\ between observation A 
and B
in all the energy bands, seems to argue against the light bending scenario.

\item

The scenario involving a mildly relativistic outflowing corona seems
to be consistent with our results. Indeed, a moderately outflowing corona 
($\beta=c/v=0.3$) can naturally
explain the relatively flat photon index observed in \3c\ (Malzac et al. 
2001). 
However, this can only be considered as circumstantial evidence in favor of 
outflowing corona models, since X-ray slopes flatter than 1.9 can  also be
produced in the framework of static corona models by 
assuming patchy configurations (Haardt et al. 1994).

\item

The jet-dominated scenario does not seem viable, as shown by previous
studies of of the time-averaged X-ray spectra of BLRGs (e.g.,
Wo\'zniak et al. 1998). Our model-independent variability results
along with the relatively large viewing angle of the jet in \3c\
(obtained from radio properties; Rudnick et al. 1986, Eracleous \&
Halpern 1998) reinforce this conclusion.  First, the fast variability
we have observed cannot be ascribed to jet emission because of the
large viewing angles (i.e., the emission is not beamed). Second, the
spectral variability (the spectral hardening when the flux decreases
and the anti-correlation of $F_{\rm var}$ with energy) is typical of
radio-quiet AGN and at odds with the spectral variability observed in
jet-dominated sources (Gliozzi et al. 2006).  Nevertheless, we cannot
exclude that an appreciable contribution from the jet component may emerge
at higher energies, as suggested by recent \suzaku\ results for the BLRG
3C~120 (Kataoka et al. 2007).

\end{itemize}

More stringent tests of the above scenarios (especially the first two)
will be afforded by more precise measurements of the black hole mass
and the bolometric luminosity (hence the Eddington ratio) of \3c. A
better mass measurement can be made using the stellar velocity
dispersion of bulge of the host galaxy (e.g., Ferrarese \& Meritt
2000; Gebhardt et~al. 2000; Tremaine et al. 2002), a technique which
leads to a small systematic error. For a better measurement of the
bolometric luminosity, the spectral energy distribution needs to be
sampled more densely, especially in the infra-red and ultra-violet
bands. Additionally, future observations with {\it Suzaku}
covering the entire X-ray range from 0.2 keV to several hundreds of
keV will be crucial for breaking the current spectral degeneracy.

Before concluding, we can speculate on the origin of the
radio-loud/radio-quiet dichotomy by putting into perspective the
results of \3c. First, our study suggests that jet production may not
necessarily be related to a very low value of the accretion rate $\dot
m$ (if the Eddington ratio is indeed in the range 0.02--0.2).
This is in agreement with findings from other BLRGs (e.g., 3C~120;
Ballantyne \& Fabian 2005, Ogle et al. 2005), radio-loud quasars
(Punsly \& Tingay 2005), and Galactic black holes (GBHs). In fact,
Galactic black holes display relativistic jets not only in the
``low-hard" state but also in the ``very-high" spectral state that is
characterized by a high accretion rate (e.g., Fender et al. 2004).
Second, a maximally rotating black hole is not a sufficient
condition for jet production (although it may be necessary).  We are
led to this conclusion by the fact that there isn't any radio jet in
MCG--6--30--15, the well known Seyfert~1 galaxy that shows the most
convincing evidence for a spinning black hole based on its
relativistically broadened \feka\ line (this galaxy harbors only a
weak, unresolved radio source; see Ulvestad \& Wilson 1984). This is
also consistent with the fact that highly relativistic jets have been
observed in neutron star binaries (e.g., Fomalont et al. 2001;
Migliari et al. 2006).  Having excluded the low $\dot m$ and the
presence of a spinning black hole, we can speculate that jet
production might be related to another parameter, such as a specific
topology of the magnetic field in the inner region of the accretion
flow. Following Livio et al. (2003),
we can hypothesize that jet production is related to formation of a
poloidal magnetic field that is triggered by an increase of the
fraction of the energy dissipated in the accretion-disk corona. This
change might cause the outflow of the corona that, in turn, can
explain the observed X-ray properties in BLRGs.

In conclusion, our study of \3c\ shows the potential 
benefits of exploiting the
complementary capabilities of \rxte\ and \chandra\, and combining
model-independent information from spectral variability with the
analysis of time-averaged X-ray spectra and morphologies. Extending
this approach to a large sample of BLRGs with higher quality data 
may help reduce the current
spectral degeneracy and shed some light in the radio-loud -quiet
dichotomy.

\begin{acknowledgements} 
We thank the anonymous referee for  constructive comments
that improved the paper.
We also thank Craig Markwardt and Keith Jahoda for their help with the issues
concerning the \rxte\ background, and Valentina Braito for enlightening
discussions. 
ME acknowledges partial support from
the Theoretical Astrophysics Visitors' Fund at Northwestern University
and thanks the members of the theoretical astrophysics group for their
warm hospitality.
\end{acknowledgements}


\begin{figure}
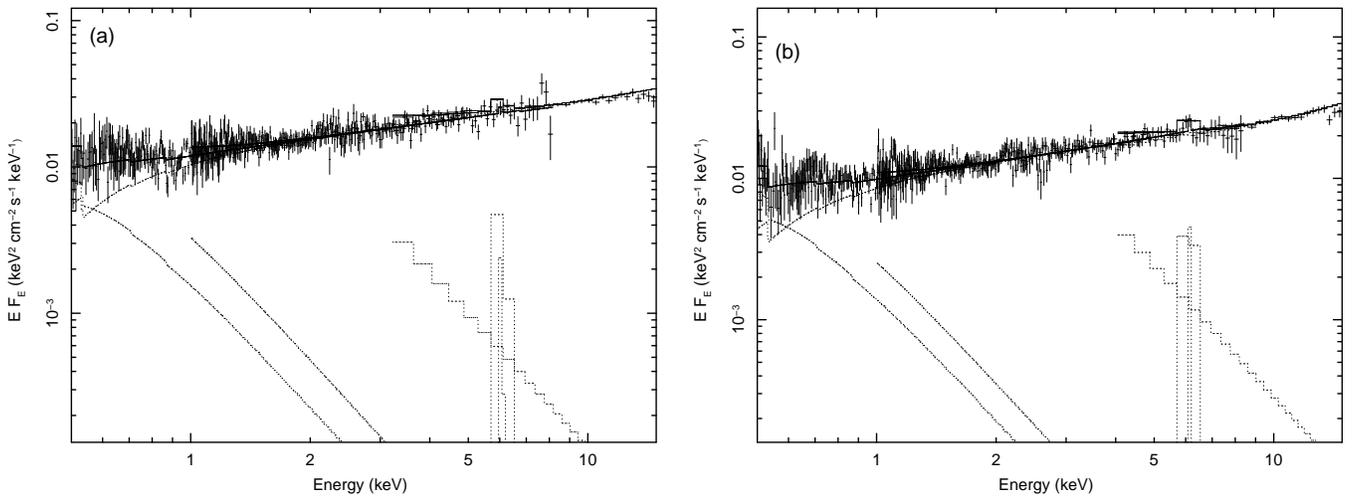

\includegraphics[bb=60 20 585 720,clip=,angle=-90,width=9.cm]{f9a.eps}
\includegraphics[bb=60 20 585 720,clip=,angle=-90,width=9.cm]{f9b.eps}
\caption{Deconvolved MEG, HEG, and PCA spectra in the 0.5--15 keV energy range,
fitted with a power-law plus a neutral reflection model (pexrav) plus
a Gaussian line at $\sim$ 6.4 keV in the source's rest frame. All models are
absorbed by
Galactic $N_{\rm H}$. Figures (a) and (b) represent observation A and B, 
respectively.
}
\label{figure:fig9}
\end{figure}

\begin{figure}
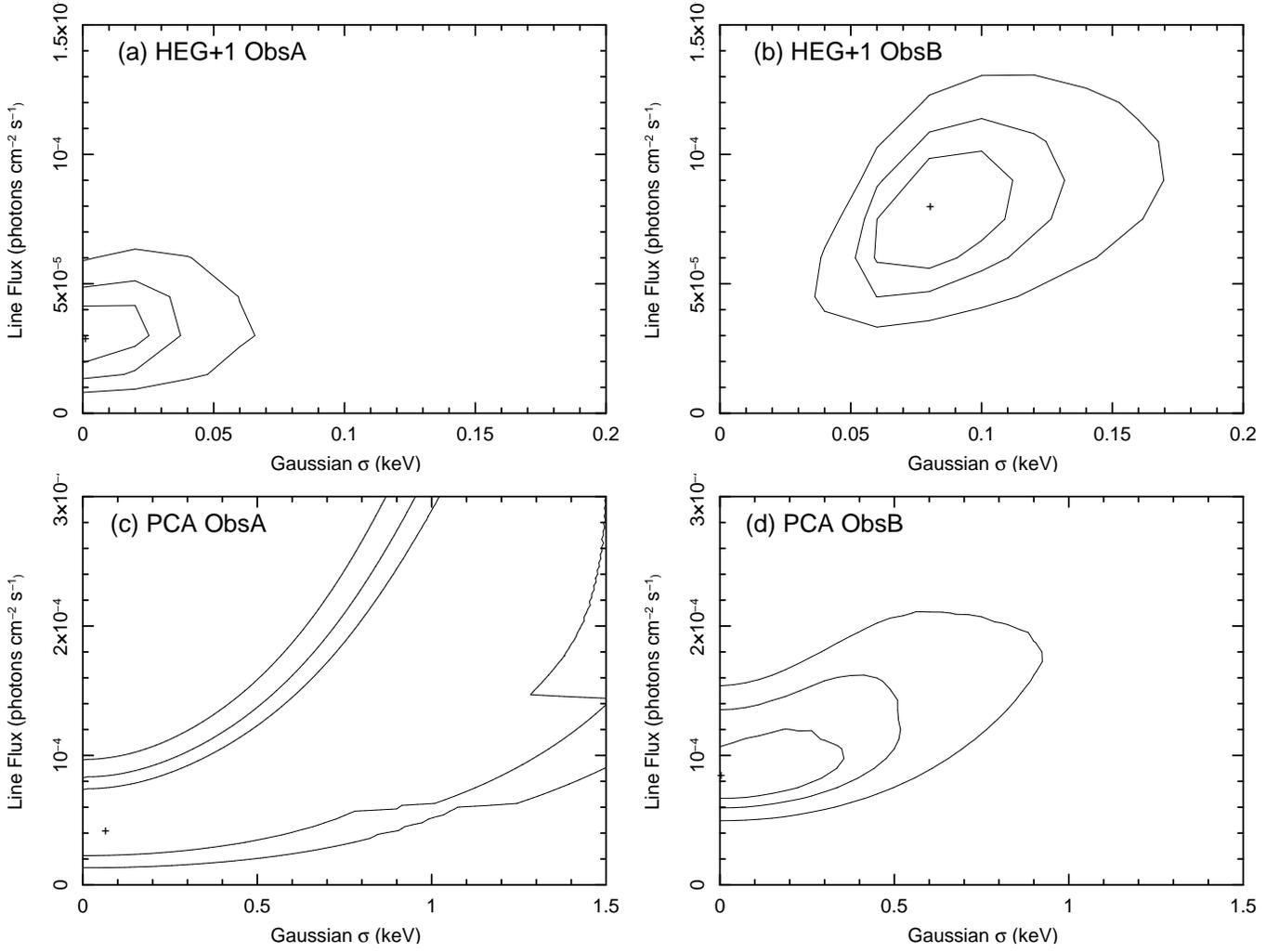

\includegraphics[bb=118 100 525 645,clip=,angle=-90,width=9.cm]{f10a.eps}
\includegraphics[bb=118 100 525 645,clip=,angle=-90,width=9.cm]{f10b.eps}

\includegraphics[bb=118 100 525 645,clip=,angle=-90,width=9.cm]{f10c.eps}
\includegraphics[bb=118 100 525 645,clip=,angle=-90,width=9.cm]{f10d.eps}

\caption{Confidence contours in the line intensity--line width
plane. The confidence levels correspond to 68\%, 90\%, and 99\%. Panel
(a) at the top left refers to observation A (HEG$+1$); panel (b) at
the top right refers to observation B (HEG$+1$). The bottom panels on
the left (c) and right (d) show the confidence contours derived from
the PCA spectra during observations A and B, respectively. }
\label{figure:fig10}
\end{figure}
    
\clearpage

\input{tab1}
\input{tab2}
\input{tab3}
\end{document}

%% file: tab1.tex
\begin{table} 
\caption{PCA Continuum Models}
\begin{center}
\begin{tabular}{cccccc} 
\hline        
\hline
\noalign{\smallskip}  
\multicolumn{6}{c}{{\bf (broken) power law}}\\
\hline 
\noalign{\smallskip}
Observation & norm & $\Gamma$ & $E_{\rm break}$ &   $\Gamma_2$    &    $\chi^2$/dof  \\ 
            & $(10^{-5})^a$   &  &   keV   &   &                   \\
\hline 
\noalign{\smallskip}
A       &$4.3_{-2.0}^{+1.9}$  &$1.78_{-.02}^{+.02}$  &   &    &  12.09/24  \\      
\noalign{\smallskip}
\hline  
\noalign{\smallskip}
B      & $8.5_{-1.9}^{+2.8}$   &$1.76_{-.03}^{+.11}$ &  $8.4_{-2.5}^{+1.1}$  & $1.53_{-.11}^{+.10}$ & 23.95/21 \\ 
\noalign{\smallskip}
\hline
\hline
\noalign{\medskip} 
\multicolumn{6}{c}{{\bf neutral reflection (pexrav) }}\\
\hline 
\noalign{\smallskip}
Observation & norm         & $\Gamma$           & R               &      &    $\chi^2$/dof  \\ 
            & $(10^{-5})^a$    &          &   $(\Omega/2\pi)$         &                        &                         \\
\hline 
\noalign{\smallskip}
A          &$4.0_{-2.0}^{+1.9}$    &$1.79_{-.05}^{+.05}$& $0.23_{-0.23}^{+0.30}$& &   10.56/23 \\      
\noalign{\smallskip}
\hline  
\noalign{\smallskip}
B        &$6.5_{-1.7}^{+1.7}$   &$1.80_{-.05}^{+.05}$&  $0.62_{-0.31}^{+0.37}$ &&   27.02/22 \\ 
\noalign{\smallskip}
\hline
\hline
\noalign{\medskip} 
\multicolumn{6}{c}{{\bf ionized reflection (pexriv) }}\\
\hline 
\noalign{\smallskip}
Observation & norm        & $\Gamma$           & R               & $\xi$ &        $\chi^2$/dof  \\ 
            & $(10^{-5})^a$    &          &   $(\Omega/2\pi)$  &    $({\rm erg~cm~s^{-1}})$                         &                         \\
\hline 
\noalign{\smallskip}
A          &$3.8_{-2.5}^{+3.6}$    &$1.80_{-.05}^{+.05}$& $0.24_{-0.24}^{+0.28}$ & $0(<5000)$ &  10.41/23 \\      
\noalign{\smallskip}
\hline  
\noalign{\smallskip}
B        &$6.6_{-2.4}^{+1.8}$   &$1.80_{-.05}^{+.06}$&  $0.60_{-0.26}^{+0.40}$ & $0(<40)$ & 26.83/21 \\ 
\noalign{\smallskip} 
\hline
\hline
\end{tabular}
\end{center}
\label{tab1}
\footnotesize
$ ^a$ Normalization of the power-law model at 1 keV in units of ${\rm~ph~cm^{-2}~s^{-1}~keV^{-1}}$.\\
A narrow ($\sigma=10$ eV) Gaussian line at 6.4 keV (in the source's rest frame) is required
in all the spectral fits. For observation B, we have also included a narrow line at 4.78 keV
(in the observer's frame) to account for the instrumental effect related to the Xe L edge.

\end{table}       

%% file: tab2.tex
\begin{table} 
\caption{MEG+HEG Continuum Models}
\begin{center}
\begin{tabular}{cccccc} 
\hline        
\hline
\noalign{\smallskip}  
\multicolumn{6}{c}{{\bf power law plus power law}}\\
\hline 
\noalign{\smallskip}
Observation & norm$_1$ & $\Gamma_{\rm soft}$ & norm$_2$ &   $\Gamma_{\rm hard}$    &    $\chi^2$/dof  \\ 
            & $(10^{-3})^a$   &  &   $(10^{-2})^a$   &   &                   \\
\hline 
\noalign{\smallskip}
A       &$2.8_{-0.7}^{+1.0}$  &$4.5_{-0.5}^{+0.5}$  &  $1.18_{-0.08}^{+0.06}$ &   $1.60_{-0.05}^{+0.04}$ &   566.3/563 \\      
\noalign{\smallskip}
\hline  
\noalign{\smallskip}
B      &$2.2_{-0.5}^{+0.6}$    &$4.6_{-0.4}^{+0.4}$ &  $0.98_{-0.05}^{+0.04}$  & $1.56_{-0.04}^{+0.03}$ & 569.8/563 \\ 
\noalign{\smallskip}
\hline
\hline
\noalign{\medskip} 
\multicolumn{6}{c}{{\bf black body plus power law }}\\
\hline 
\noalign{\smallskip}
Observation & norm$_1$         & $kT$           & norm$_2$      & $\Gamma$               &          $\chi^2$/dof  \\ 
            & $(10^{-4})^b$    & (keV)          &   $(10^{-2})^a$     &                        &                         \\
\hline 
\noalign{\smallskip}
A          &$3.5_{-0.6}^{+0.8}$    &$0.10_{-0.01}^{+0.01}$& $1.31_{-0.02}^{+0.02}$ & $1.69_{-0.02}^{+0.02}$&   577.6/564 \\      
\noalign{\smallskip}
\hline  
\noalign{\smallskip}
B        &$2.9_{-0.5}^{+0.7}$   &$0.10_{-0.01}^{+0.01}$&  $1.07_{-0.02}^{+0.02}$ & $1.64_{-0.02}^{+0.02}$&   575.4/564 \\ 
\noalign{\smallskip}
\hline
\hline
\noalign{\medskip} 
\multicolumn{6}{c}{{\bf bremsstrahlung plus power law  }}\\
\hline 
\noalign{\smallskip}
Observation & norm$_1$  & $kT$          & norm$_2$    & $\Gamma$                 &    $\chi^2$/dof  \\ 
            & $(10^{-1})^c$   & (keV)   &   $(10^{-2})^a$         &                          &                \\
\hline 
\noalign{\smallskip}
A        &$1.3_{-0.2}^{+0.2}$    &$0.19_{-0.02}^{+0.03}$& $1.29_{-0.02}^{+0.02}$ &$1.68_{-0.02}^{+0.02}$&    579.2/564 \\      
\noalign{\smallskip}
\hline  
\noalign{\smallskip}
B        & $1.0_{-0.3}^{+0.5}$  & $0.20_{-0.02}^{+0.03}$ & $1.06_{-0.02}^{+0.02}$ & $1.63_{-0.02}^{+0.02}$&     574.4/564 \\ 
\noalign{\smallskip}
\hline
\hline

\end{tabular}
\end{center}
\label{tab2}
\footnotesize
$ ^a$ Normalization of the power-law model at 1 keV in units of ${\rm~ph~cm^{-2}~s^{-1}~keV^{-1}}$.
$^b$ Normalization of the black body model at in units of $L_{39}/D^2_{10}$, where $L_{39}$ is the
source luminosity in ${\rm 10^{39}~erg~s^{-1}}$ and $D_{10}$ the distance of the source in units of
10 kpc.
$^c$ Normalization of the Bremsstrahlung model; see the Xspec manual for a detailed description.

\end{table}       

%% file: tab3.tex
\begin{table} 
\caption{Fe K$\alpha$ Gaussian Model Parameters (Rest Frame)}
\begin{center}
\begin{tabular}{clclccc} 
\hline        
\hline
\noalign{\smallskip} 
Observation & Instrument & $E_{\rm Fe}$  &$\sigma_{\rm Fe}$ &  $I_{\rm Fe}$                       & EW    &    $\Delta\chi^2$/dof  \\ 
            &            &  (keV)        &   (eV)           &  ($10^{-5}{\rm~ph~cm^{-2}~s^{-1}}$) & (eV)  &    ($\Delta C$/dof)  \\
\hline 
\noalign{\smallskip}
A           &PCA&  $6.42_{-0.30}^{+0.30}$&3 ($<$955)      & $4.2_{-2.0}^{+2.0}$&$58_{-28}^{+28}$&  -13.0/3\\      
\noalign{\smallskip}
\hline
\noalign{\smallskip}
A           &HEG+1-1&  $6.37_{-0.05}^{+0.03}$ & 1($<$94)     & $1.2_{-1.0}^{+1.4}$ & $17_{-15}^{+11}$ &  -3.7/3\\      
\noalign{\smallskip}
A           &HEG+1&    $6.62_{-0.14}^{+0.01}$ &  2($<$30)      & $2.9_{-1.3}^{+1.5}$&$47_{-22}^{+24}$&  -16.1/3 \\
\noalign{\smallskip}
\hline
\hline
\noalign{\smallskip}
B           &PCA&  $6.50_{-0.16}^{+0.10}$&5 ($<$356)      & $8.3_{-2.0}^{+16}$&$133_{-30}^{+255}$&  -54.0/3\\      
\noalign{\smallskip}
\hline
\noalign{\smallskip}
B           &HEG+1-1&  $6.43_{-0.07}^{+0.05}$ &  6($<$87)    & $3.5_{-1.3}^{+2.8}$&$55_{-20}^{+47}$&  -10.4/3\\      
\noalign{\smallskip}
B           &HEG+1& $6.39_{-0.04}^{+0.03}$      &  $81_{-25}^{+35}$  & $7.3_{-2.2}^{+3.2}$&$132_{-46}^{+42}$&  -35.2/3 \\
\noalign{\smallskip}
\hline

\hline

\end{tabular}
\end{center}
\label{tab1}
\end{table}       